\documentstyle[prl,aps,graphics,epsfig]{revtex}

\input{psfig.sty}
\begin{document}
\pagestyle{empty}
\draft
\twocolumn
\wideabs{
\title{Daylight quantum key distribution over 1.6 km}
\author{W.T. Buttler, R.J. Hughes, S.K. Lamoreaux, G.L. Morgan, J.E. Nordholt 
and C.G. Peterson}
\address{University of California, Los Alamos National Laboratory, Los Alamos, 
New Mexico 87545}
\date{\today}
\maketitle
\begin{abstract}
Quantum key distribution (QKD) has been demonstrated over a point-to-point 
$\sim1.6$-km atmospheric optical path in full daylight. This record transmission 
distance brings QKD a step closer to surface-to-satellite and other 
long-distance applications.
\end{abstract}
\pacs{PACS Numbers: 03.65.Bz, 42.79.Sz}
}
\narrowtext
Quantum cryptography was introduced in the mid-1980s \cite{ref:BB84} as a new 
method for generating the shared, secret random number sequences, known as 
cryptographic keys, that are used in crypto-systems to provide communications 
security (for a review see \cite{ref:ContPhys}). The appeal of quantum 
cryptography (or more accurately, quantum key distribution, QKD) is that its 
security is based on laws of nature and information-theoretically secure 
techniques, in contrast to existing methods of key distribution that derive 
their security from the perceived intractability of certain problems in number 
theory, or from the physical security of the distribution process.

Several groups have demonstrated QKD over multi-kilometer distances of optical 
fiber \cite{ref:fiber}, but there are many key distribution problems for which 
QKD over line-of-sight atmospheric paths would be advantageous (for example, it 
is impractical to send a courier to a satellite). Free-space QKD was first 
demonstrated in 1990 \cite{ref:32-cm,ref:privamp2} over a point-to-point 32-cm 
table top optical path, and recent work has produced atmospheric transmission 
distances of 75 m \cite{ref:FreeSpace} (daytime) and 1 km \cite{ref:Buttler2} 
(nighttime) over outdoor folded paths (to a mirror and back). The close 
collocation of the QKD transmitter and receiver in folded-path experiments is 
not representative of practical applications and can result in some compensation 
of turbulence effects. We have recently performed the first point-to-point 
atmospheric QKD in full daylight, achieving a 0.5-km transmission range 
\cite{ref:hughes-1/2-km}, and here we report a record 1.6-km point-to-point 
transmission in daylight, with a novel QKD system that has no active 
polarization switching elements.

The success of QKD over atmospheric optical paths depends on the transmission 
and detection of single-photons against a high background through a turbulent 
medium. Although this problem is difficult, a combination of temporal, spectral 
\cite{ref:PhotByPhot,ref:DayPairs} and spatial filtering \cite{ref:Buttler} can 
render the transmission and detection problems tractable 
\cite{ref:hughes-1/2-km}. The essentially non-birefringent nature of the 
atmosphere at optical wavelengths allows the faithful transmission of the 
single-photon polarization states used in the free-space QKD protocol.

A QKD procedure starts with the sender, ``Alice,'' generating a secret random 
binary number sequence. For each bit in the sequence, Alice prepares and 
transmits a single photon to the recipient, ``Bob,'' who measures each arriving 
photon and attempts to identify the bit value Alice has transmitted. Alice's 
photon state preparations and Bob's measurements are chosen from sets of 
non-orthogonal possibilities. For example, using the B92 protocol \cite{ref:B92} 
Alice agrees with Bob (through public discussion) that she will transmit a 
$45^\circ$ polarized photon state $| 45 \rangle$, for each ``0'' in her 
sequence, and a vertical polarized photon state $| v \rangle$, for each ``1'' in 
her sequence. Bob agrees with Alice to randomly test the polarization of each 
arriving photon with $-45^\circ$ polarization, $| \! -45 \rangle$, to reveal 
``1s,'' or horizontal polarization, $| h \rangle$, to reveal ``0s.'' In this 
scheme Bob will never detect a photon for which he and Alice have used a 
preparation/measurement pair that corresponds to different bit values, such as 
$|h\rangle$ and $|v\rangle$, which happens for 50\% of the bits in Alice's 
sequence. However, for the other 50\% of Alice's bits the preparation and 
measurement protocol uses non-orthogonal states, such as for $| 45 \rangle$ and 
$|h\rangle$, resulting in a 50\% detection probability for Bob. Thus, by 
detecting single-photons Bob identifies a random 25\% portion of the bits in 
Alice's random bit sequence, assuming a single-photon Fock state with no bit 
loss in transmission or detection. This 25\% efficiency factor, $\eta_Q$, is the 
price that Alice and Bob must pay for secrecy.

Bob and Alice reconcile their common bits by revealing the locations, but not 
the bit values, in the sequence where Bob detected photons; Alice retains only 
those detected bits from her initial sequence. In practical systems the 
resulting sifted key sequences \cite{ref:N-Lutk}, will contain errors; a pure 
key is distilled from them using classical error detection techniques. The 
single-photon nature of the transmissions ensures that an eavesdropper, ``Eve,'' 
can neither ``tap'' the key transmissions with a beam splitter (BS), owing to 
the indivisibility of a photon \cite{ref:indivis}, nor faithfully copy them, 
owing to the quantum ``no-cloning'' theorem \cite{ref:noclone}. Furthermore, the 
non-orthogonal nature of the quantum states ensures that if Eve makes her own 
measurements she will be detected through the elevated error rate she causes by 
the irreversible ``collapse of the wavefunction'' \cite{ref:quantum-colapse}. 
From the observed error rate and a model for Eve's eavesdropping strategy,
Alice and Bob can calculate a rigorous upper bound on the infomation Eve might 
have obtained. Then, using the technique of generalized privacy amplification by 
public discussion \cite{ref:PrivAmp} Alice and Bob can distill a shorter, final 
key on which Eve has less than one bit of information.

\begin{figure}[!h]
\psfig{figure=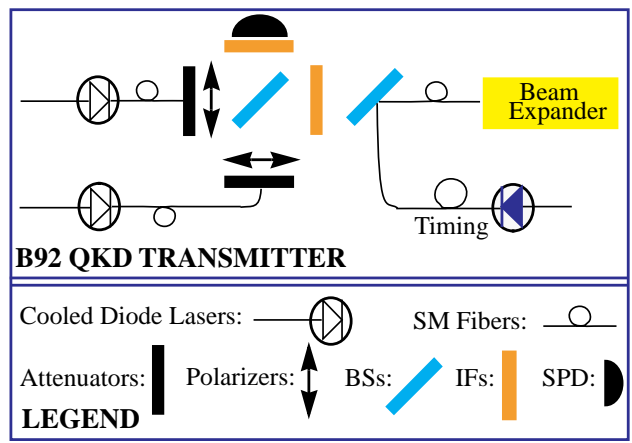,width=3.25 in}
\caption{Free-Space QKD Transmitter (Alice).}
\label{fig:qkd-trans}
\end{figure}

The QKD transmitter (``Alice'') in our experiment (Fig. 
\ref{fig:qkd-trans})operates at a clock rate $R_0 = 1$-MHz. On each ``tick'' of 
the clock a $\sim 1$-ns vertically-polarized optical ``bright pulse''
is produced from a ``timing-pulse'' diode laser whose wavelength is temperature 
controlled to $\sim 768$ nm. After a $\sim 100$-ns delay one of two 
temperature-controlled dim pulse ``data'' diode lasers emits a $\sim 1$-ns 
optical pulse that is attenuated to the single-photon level \cite{ref:Note3} and
constrained by an interference filter (IF) to $773 \pm 0.5$ nm to remove 
wavelength information. Polarizers set one data laser's output to be $45^\circ$ 
polarized and the other to be vertically polarized as required for the B92 
protocol. The choice of which data laser fires is determined by a random bit 
value that is obtained by discriminating electrical noise. The random bit value 
is indexed by the clock tick and recorded in Alice's computer control system's 
memory. All three optical pulse paths are combined with beamsplitters (BSs) into 
a single-mode (SM) optical fiber to remove spatial mode information, and 
transmitted toward Bob's receiver through a $27 \times$ beam expander (to extend 
the system Rayleigh range). A single-photon detector (SPD) \cite{ref:SPD} 
located behind a matched IF in one of the BS output ports is used to monitor the 
average photon number $\bar{n}$ of the dim-pulses as follows: (1) a calibration 
photon-number measurement is made from the rate at which a calibrated 
single-photon counting module (SPCM) \cite{ref:SPCM} fires at the transmitter's 
SM transmission-fiber output with a given input, (2) next the transmitter's SPD 
count rate is calibrated to the SPCM firing rate with the same input to 
determine the SPD efficiency, which is then (3) used with the experimental SPD 
count rates to measure the transmitted $\bar{n}$ in key generation mode.

\begin{figure}[!h]
\psfig{figure=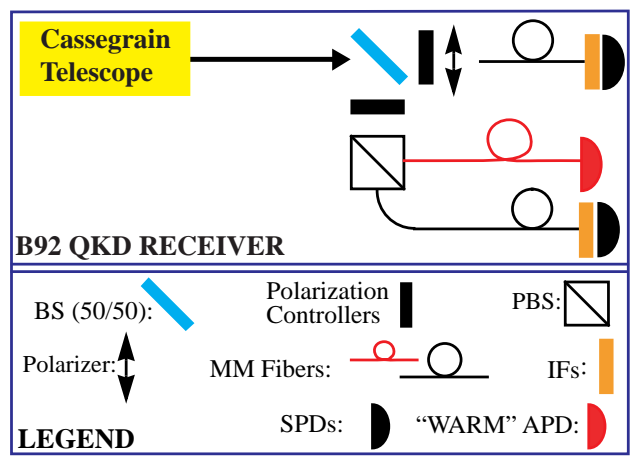,width=3.25 in}
\caption{Free-Space QKD Receiver (Bob).}
\label{fig:qkd-rec}
\end{figure}

At the QKD receiver (``Bob'') light pulses are collected by a $8.9$-cm diameter
Cassegrain telescope and directed into a polarization analysis and detection 
system (Fig. \ref{fig:qkd-rec}). A bright pulse triggers a ``warm'' avalanche 
photodiode (APD), which sets up a narrow $\sim 5$-ns coincidence gate in which 
to test a subsequent dim pulse's polarization \cite{ref:Note1}. A BS randomly 
directs dim pulses along one of two paths. Polarization elements along the upper 
path are set to transmit $-45^\circ$ polarization in accordance with Bob's B92 
``1'' value, while along the lower path a measurement for $| h \rangle$ to 
reveal ``0''s is made using a polarizing beamsplitter (PBS). (The PBS transmits 
$| h \rangle$ but reflects $| v \rangle$.) Each analysis path contains a matched 
IF and couples to a SPD via multi-mode (MM) fiber that provides limited spatial 
filtering, giving the receiver a restricted $200$ $\mu$-radian field of view. 
For events on which one of the two SPDs triggers during the coincidence gate, 
Bob can assign a bit value to Alice's transmitted bit; upper-path SPD firings 
identify ``1''s, and lower-path SPD firings identify ``0''s. He records these 
detected bits in the memory of his computer control system, indexed by the 
``bright pulse'' clock tick. Bit generation is completed when Bob communicates 
the locations, but not values, of his photon detections in Alice's random 
bit-sequence over a public channel: wireless ethernet in our experiment.

The QKD system was operated over a $1.6$-km outdoor range with excellent 
atmospheric conditions on Friday $13$ August $1999$ beginning at 09:30 LST under 
cloudless New Mexico skies. By 11:30 LST turbulence induced beam-spreading 
hindered our ability to efficiently acquire data at low bit-error rates ($BER$), 
$\epsilon$ (where $BER$, $\epsilon$, is defined as the ratio of the number of 
bits received in error to the total number of bits received). The system 
efficiency, $\eta_{system}$, which accounts for losses between the transmitter 
and MM fibers at the receiver, and the receiver's SPDs efficiencies had an 
average value of $\langle \eta_{system} \rangle \sim 0.13$ with a standard 
deviation of $\sigma = 0.04$. Fluctuations in $\eta_{system}$ were caused by 
turbulence induced beam-spreading and beam-wander; the typical beam-wander was 
observed to be on the order of $3$ to $5$ $\mu$-radians. (Our present system has 
no beam-steering or adaptive-optics technology to compensate for 
turbulence-induced effects.) The $\eta_Q = 0.25$ quantum efficiency of the B92 
protocol lowers the overall efficiency to $\eta = \eta_Q \eta_{system} \sim 
0.0325$ and leads to a detection probability for Bob of $P_B = 1 - exp(- \eta \, 
\bar{n})$. This gave a bit-rate of $R \sim 5.4$, $12.2$, and $17$-kHz at 
$\bar{n} \sim 0.2$, $0.35$, and $0.5$-photons per dim-pulse, respectively, when 
the lasers were pulsed at $R_0 = 1$ MHz. Bits were transmitted in $25$, $50$, 
and $100$ k-bit blocks. A total of $1.55$ M-bits were sent in $40$ data 
exchanges between Alice and Bob and $17,420$ bits of sifted key were received. 
Table \ref{table:key_bits} includes a typical $250$-bit sample from one of 
several $1.6$-km daylight transmissions on $13$ August $1999$. The sifted key 
shown contains eight bit-errors (in bold) corresponding to $\epsilon = 3.2$\% 
for these $250$ bits and has a $60$:$40$ bias toward ones (the average bias for 
all experiments on $13$ August $1999$ was $50.3$:$49.7$ toward ones). The 
average $BER$ on all key material acquired during the daylight transmissions was 
$\langle \epsilon \rangle = 5.3$\%. These $BER$s would be regarded as 
unacceptably high in any conventional telecommunications application but are 
tolerated in QKD because of the secrecy of the bits.

The dominant $BER$ component is from the ambient solar background, with a 
measured noise probability for both detectors of about $6.7 \times 10^{-4}$ per 
coincidence gate, contributing about $5.9$\% to the $\langle \epsilon \rangle = 
7.8$\% at $\bar{n} = 0.2$ data, about $2.4$\% to the $\langle \epsilon \rangle = 
4.1$\% at $\bar{n} = 0.35$, and about $1.9$\% to the $\langle \epsilon \rangle = 
4.1$\% at $\bar{n} = 0.5$. (The ambient-background is somewhat less than that 
expected from the daylight radiance \cite{ref:Buttler2}, which we attribute to 
Bob viewing the dark interior of the tent housing Alice's transmitter.) 
Imperfections and misalignments of the polarizing elements were the next largest 
contribution (about $1.9$\%) to the total $BER$s on 13 August 1999. Experience 
from previous experiments \cite{ref:hughes-1/2-km,ref:Buttler2,ref:Buttler} 
suggests that this component of $BER$ can be reduced to about $0.5$\%. Detector 
dark noise ($\sim 1,400$ dark-counts per second) makes an even smaller 
contribution of $< 0.1$\% to the $BER$. The dual-fire rate --- the probability 
that both SPDs fire during a coincidence window--- was $0.0003$, $0.0007$, and 
$0.001$ at $\bar{n} \sim 0.2$, $0.35$, and $0.5$, respectively.

Alice and Bob can correct errors by transmitting error correction information 
over the public channel, amounting to
\begin{equation}
   f(\epsilon) = - \epsilon \log_2 \epsilon  - (1 - \epsilon) \log_2
                (1 - \epsilon)
   \label{eq:shannon_limit}
\end{equation}
bits per bit of sifted key in the Shannon limit. For example, for $\epsilon = 
4.1$\%, $f(0.041) = 0.246$. Practical error-correcting codes do not achieve the 
Shannon limit, although the interactive scheme known as {\it Cascade} 
\cite{ref:cascade}, comes within about $1.16 \: f(\epsilon)$ for error rates up 
to $5$\% \cite{ref:N-Lutk}. Our experiments use a combination of block-parity 
checks and Hamming codes \cite{ref:hamming} achieving an efficiency equivalent 
to the $Cascade$ scheme but with greater computational efficiency. The error 
correction information is transmitted over the public channel and thus could 
provide information about the key material to Eve, reducing Alice and Bob's 
secret bit yield. (Alice and Bob could encrypt the error correction information 
to deny Eve access to it, but at the cost of an equal number of shared secret 
key bits \cite{ref:sig-auth1}.)

Alice and Bob now use ``privacy amplification'' \cite{ref:PrivAmp} to reduce any 
partial knowledge gained by an eavesdropper to less than 1-bit of information. 
(For discussions of eavesdropping strategies see References 
\cite{ref:N-Lutk,ref:durt,ref:buttler3}.) We have not implemented privacy 
amplification at this time, but to estimate the secret-key rate for our 
experiment and its dependencies on relative parameters, we assume Eve is 
restricted to performing the combination of the intercept-resend and 
beamsplitting attacks considered in \cite{ref:32-cm,ref:privamp2}. In this case 
Alice and Bob could use the parities of random subsequences of their 
error-corrected keys as their final secret key bits, resulting in a compression 
to
\begin{equation}
   F(\epsilon) = (1 - \bar{n}) - 2 \: \sqrt{2} \: \epsilon
   \label{eq:optimal_nbar}
\end{equation}
bits per bit of error-corrected key. The first term in Eq. \ref{eq:optimal_nbar} 
accounts for the multi-photon fraction of Alice's dim-pulses, which are 
susceptible to beamsplitting, while the second accounts for Eve performing 
intercept-resend on a fraction of the pulses. The final secret bit yield is 
therefore a fraction $F(\epsilon)  - f(\epsilon)$ the length of the original 
sifted key. For $\bar{n} \lesssim 0.05$, under the conditions of our $13$ August 
$1999$ experiment with $\eta_{system} = 0.13$, there is no net secret bit yield
because of the large value $f(\epsilon)$. With increasing $\bar{n}$ the $BER$ 
decreases so rapidly that the increased privacy amplification cost to protect 
against beamsplitting is more than offset by the reduced error-correction cost, 
and so the secret bit yield initially increases. However, for larger $\bar{n}$ 
values, the privacy amplification factor $F(\epsilon)$ required to compensate 
for beamsplitting of multi-photon pulses becomes small, and the secret bit yield 
decreases. For $\eta_{system} = 0.13$, we find that the optimum $\bar{n}$ for 
our $13$ August $1999$ experiment is $\sim 0.4$ giving a secret bit yield of 
$38.5$\% of the sifted key length, and $\sim 0.4$\% of the length of the 
transmitted sequence. (With $Cascade$ the optimal $\bar{n}$ would also be $\sim 
0.4$ and the secret bit yield would be $24.7$\% of the sifted key or $0.32$\% 
the length of the transmitted sequence, giving a secret bit rate of $\sim 
3$-kHz.) For smaller $\eta_{system}$ values under $13$ August $1999$ conditions 
the optimal $\bar{n}$ values are as above but the secret bit yield is smaller; 
for $\eta_{system} < 0.04$ there is no secret bit yield. (To protect against the 
attacks proposed in \cite{ref:N-Lutk}, should they become feasible, we would 
need to reduce our background further with a shorter coincidence gate window and 
narrower spectral filters to have a non-zero secret-bit yield at the $\bar{n}$ 
values required.)

This paper reports QKD between a transmitter and receiver separated by a 
$1.6$-km daylight atmospheric optical path. This transmission distance, which 
was only limited by the length of the available range, is the longest to date. 
Our system has no active polarization elements, resulting in greater simplicity 
and security over previous experiments. Secret bit rates of several kilo-Hertz 
protected against simple beamsplitting and intersept-resend attacks have been 
shown to be feasible. Such rates would enable the rekeying of cryptographic 
systems. The system could be easily adapted to the BB84 four-state QKD protocol 
\cite{ref:BB84} or to use single-photon light sources \cite{ref:down-conversion} 
once they are available, providing protection against more sophisticated future 
attacks \cite{ref:N-Lutk,ref:durt,ref:buttler3}. Our results are representative 
of practical situations showing that QKD could be used in conjunction with 
optical communication systems and providing further evidence for the feasibility 
of surface-to-satellite QKD \cite{ref:hughes-1/2-km}. The $1.6$-km optical path 
is similar in optical depth to the effective turbulent atmospheric thickness 
encountered in a surface-to-satellite application. Significant amounts of 
key-material (about $15$ k-bits) with low $BER$s ($\langle \epsilon \rangle 
\lesssim 3.0$\%) at low $\bar{n}$ ($\bar{n} \lesssim 0.2$) were also taken at 
night and during light rain over this $1.6$-km distance. Finally, we note that 
the variability of system efficiency and background is a feature of atmospheric 
QKD that is quite different from optical fiber systems.
We wish to thank P. G. Kwiat for QKD discussions, D. Derkacs for technical
support, and G. H. Nickel, D. Simpson, and E. Twyeffort for discussions
regarding bit-error detection protocols and privacy amplification.
%
%
\vspace{-1.25cm}
%
\vspace{-0.5cm}
\begin{table}[!h]
\caption{A 250-bit sample of Alice's (a) and Bob's (b) raw key material
generated at Los Alamos, New Mexico at 10:00 LST (GMT $-$ $7$) on Friday
$13$ August $1999$. Alice was located at $1978$-m elevation, $35^\circ \:
46.859^\prime$ N, and $106^\circ \: 14.932^\prime$ W; Bob was located at
$1966$-m elevation, $35^\circ \: 46.376^\prime$ N, and $106^\circ \:
14.052^\prime$ W. The beam height at Alice's transmitter and Bob' receiver
was $1.5$-m; the maximum beam height of $107$-m above the terrain occurred
$1$ km from Alice; and the average beam height above the terrain was $\sim
38$-m.}
\begin{center}
\begin{tabular}{|c|c|}
a & 00011011110111010111010000101011111101111101110000 \\
b & {\bf 1}0011011110{\bf 0}1101011{\bf 0}01{\bf
1}000101011111101111101110000
\\ \hline
a & 01111110111100011011000010111101110010000101001010 \\
b & 011111101{\bf 0}11000110110000{\bf 0}0111101110010000101001010 \\ \hline
a & 00011110111110000100011111001111011011011101101111 \\
b & 0001111011{\bf 0}110000100011111001111011011011101101111 \\ \hline
a & 10010010100100100100111100000001101001111100101111 \\
b & 10010010100100100100111100000001101001111100101{\bf 0}11 \\ \hline
a & 11111111111111111000011111011101101110101100011101 \\
b & 11111111111111111000011111011101101110101100011101 \\
\end{tabular}
\end{center}
\label{table:key_bits}
\end{table}

\begin{thebibliography}{99}
\bibitem{ref:SPD} Our SPDs are based on EG\&G Ca\-na\-da Ltd., Optoelectronics 
Division, product \#C30902S, and have a single photon detection efficiency of 
$\eta_{spd} \sim 0.69$ at $773$ nm, and a dark count rate $\sim 1,400$ s$^{-1}$.
\bibitem{ref:SPCM} The MM fiber-coupled SPCM is a EG\&G Ca\-na\-da Ltd., 
Optoelectronics Division, product and has a single photon detection
efficiency of $\eta_{spcm} \sim 0.44$ at $773$ nm, and a dark count rate $< 100$ 
s$^{-1}$.
\end{thebibliography}
\end{document}